# Plasmonic nanowire coupled to zero-dimensional nanostructures: A brief review

Sunny Tiwari[1], Chetna Taneja[1] and G V Pavan Kumar[1, 2 *]

[1]*Department of Physics, Indian Institute of Science Education and Research, Pune-411 008, India*
[2]*Center for Energy Science, Indian Institute of Science Education and Research, Pune-411 008, India*

Metal nanowires and nanoparticles that facilitate surface plasmons are of contemporary interest in nanophotonics, thermoplasmonics and optoelectronics. They facilitate not only subwavelength light propagation and localization capabilities, but also provide an excellent platform for opto-thermal effects confined to volumes down to the nanoscale. This brief review article aims to provide an overview of a specific nanophotonic geometry: a plasmonic nanowire coupled to a zero-dimensional nanostructure. We discuss the methods to prepare such nano-architectures and review some interesting nanophotonic applications that arise out of it. We conclude with a discussion on some emerging research directions that can be facilitated by employing the coupled nanostructures.

## 1 Introduction

Propagation, localization and routing of light at subwavelength scales have been an important task in nanophotonics [1-3]. The relevance of this task is motivated by the fact that optical and optoelectronic devices are to be shrunk to smaller scale with minimal footprint, and to this end, efforts have been made to address pertinent questions such as: what is the efficient way to route light from one location to another on a nano-chip? What are the limiting factors that hinder photonic information transfer? What are the materials that can be utilized for on-chip optical information processing? What new physical phenomenon can emerge from the interaction of light with matter at nanoscale? As one can observe, the boundaries between fundamental questions and applications are no more distinct, and this facilitates an interesting opportunity for researchers to look into problems that are relevant to optical physics and optical engineering.

An important aspect of surface plasmon polaritons (SPPs) facilitating nanomaterials is the dimensionality of the nanostructure [4]. A variety of nanostructures have been realized using both top-down and bottom-up nanofabrication strategies [5-7]. Quasi-zero dimensional metal nanoparticles of various shapes and sizes have been realized, and these structures, especially in the coupled form, are excellent candidates to localize light as plasmonic resonators [8]. Thanks to the extremely small mode volumes of such localized plasmon resonators, one can confine a large magnitude of electric field at gaps of metal nanostructures, and thus can be harnessed for single molecule sensing and low-power nonlinear optics [9-14].

Quasi-one dimensional metal nanostructures in the form of nanowires and nanobelts facilitate delocalized SPPs [4]. This means one can propagate an optical signal from one location to another via such structures, thus acting as a waveguide [15]. Unlike conventional optical fibers, plasmonic nanowire are not constrained by the diffraction limit of light and hence can be employed as sub-wavelength optical waveguides. This ability to couple and route light beyond diffraction limit has created an impetus to explore emergent physics and applications that can arise out of this quasi-one dimensional nanostructures [16,17].

An interesting prospect that is of relevance to nanophotonics is the coupling between quasi-zero dimensional (0-D) nanostructures (e.g. metal nanoparticle) with quasi-one dimensional (1-D) nanostructures (e.g. metal nanowires). Such coupling can lead to some optical properties that are otherwise not present in the individual nanostructures taken in isolation. In this review article, we aim to give a brief summary of one such coupling between 1-D and 0-D nanostructures e.g: a plasmonic nanowire coupled to a zero- dimensional nanostructures. We will give a glimpse of how such nanostructures are fabricated, and how they can be utilized to study light-matter interaction at sub-wavelength scales. Towards the end, we will briefly summarize some emerging opportunities of the coupling between 1-D and 0-D structures.

## 2 Nanowire as a waveguide

Metal nanostructures in the form of nanowires (NWs) facilitate delocalized SPPs and thus can be used to propagate optical signals in the sub-diffraction limit, from one location to another by acting as a waveguide. NWs prepared chemically are atomically smooth and thus suffer lower losses as compared to NWs prepared with electron beam lithography or any other top down technique. Gold nanowires (AuNWs), when prepared by bottom up approach are generally thin and good resonators but suffer absorption losses in the lower part of visible spectrum [18]. Tuning the length and size of these NWs is also a difficult task. On the other hand, silver nanowires (AgNWs) with different thicknesses can be prepared by changing the parameters of the chemical synthesis. A chemically prepared, single crystalline, few hundreds of nanometers thick AgNWs are better waveguides which suffer minimal losses of optical signal during propagation. The most common and efficient method to prepare single crystalline silver nanowire is the polyol process [19]. Silver nitrate is reduced into silver by heating in ethylene glycol and polyvinyl pyrrolidone (PVP) solution. Ethylene glycol and PVP both act as reducing and capping agents for the formation of NWs from reduced silver. Depending on the ratio of silver nitrate and PVP concentration and the heating time, the length and thickness of nanowire can be tuned.

SPPs along the NWs can be excited in various configurations such as by focusing laser at one end, evanescent field produced by total internal reflection, near field coupling of emission by molecules placed in the vicinity of nanowire [20,21]. Figure (1a) shows the schematic of SPPs excitation along AgNW [20]. End I of AgNW is excited by focused incoming laser with polarization along the NW. As shown in Fig (1b) plasmons along the NW are excited by tightly focusing the light at one of the ends. This end excitation acts as a defect and scatters the light in all directions. Some of the scattered photons with larger angles have sufficient momentum to excite the plasmons along the NW. The SPPs after propagating along the NW propagate to the distal end and outcouples as freely propagating photons as shown by the red arrow. At the distal end, the plasmons can also be reflected back and thus both the ends of AgNW can confine the plasmons propagation along the NW making it a Febry-Pérot type oscillator [20,22]. Figure (1c) shows the scanning near-field optical microscopy (SNOM) image of near field electric field profile of SPPs along a small part (white box in Fig 1b) of NW. The modulation of near-field electric field of SPPs shows the Febry-Pérot nature and is represented in inset (i) of Fig (1c). The modulation of intensity as a function of wavelength is shown in Fig (1d) and thus confirms the Febry-Pérot resonance nature of propagating SPPs along the NW. Since the chemically prepared NWs are single crystalline and thus they act as a better reflector for SPPs traveling along the NWs. A lithographically prepared AgNW does not show large modulation because of poor reflection of SPPs from the NW ends.

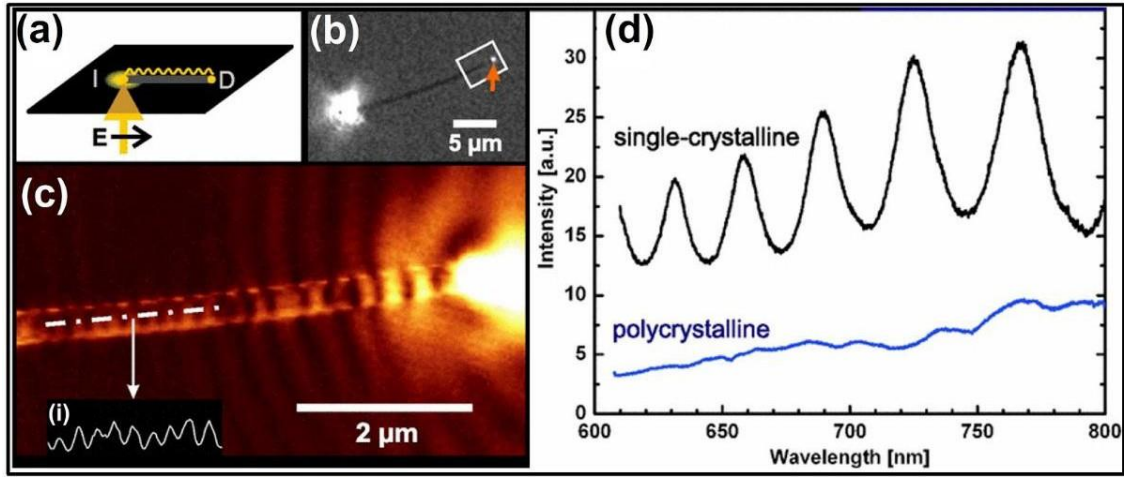

Fig 1. Waveguiding property of single crystalline silver nanowire showing propagation of surface plasmonpolaritons. (a) Schematic showing the surface plasmon excitation at input end I and outcoupling at distal end D. The polarization is kept along the length of nanowire. (b) Optical image of silver nanowire. White rectangular box shows emission at the distal end of nanowire when exciting the other end. (c) Scanning near field optical microscopy image of the area under white rectangular box in Fig (1b). Inset (i): Intensity cross-cut along the white dotted line on silver nanowire in Fig (1c). (d) Spectra collected from the distal end of a chemically prepared single-crystalline silver nanowire (black curve) and a lithographically prepared silver nanowire (blue curve). Reproduced with permission from [20].

**3 Coupling 0-dimensional structure to 1-dimensional structure**

Optical fields that are delocalized but confined to sub-wavelength scales are of relevance in on-chip optical processing. Coupling light into and out of such systems needs a specific location on the geometry. Zero-dimensional nanostructures such as nanoparticles and quantum dots can serve this purpose [23,24]. We discuss various fabrication procedures to realize these structures.

*(i) Capillary force based self-assembly*

Capillary forces are known to act as a driving force in various self-assembly of nanostructures [25, 26]. This force can be used to assemble any plasmonic structure to another structure to create plasmonic junction or a hotspot which is a region of high electric field. Figure (2a) shows one such schematic to understand and calculate this capillary force which has been used to assemble a single gold nanoparticle [27,28]. During the drying process of Au colloids near silver nanowires the capillary force provides a long range interaction between nanowire-nanoparticle (NW-NP). The force calculated for a ~ 90 nm AuNP and ~ 100 nm AgNW is $2.26 \times 10^{-8}$ N to $1.13 \times 10^{-8}$ N which can easily overcome the effect of gravity on AuNP [29]. Thus, the capillary force of absorption from NW on NP (imbibition) overcomes the gravity force on single AuNP and the particle goes and assemble near AgNW forming a hotspot or NW-NP junction. But depending on the concentration of NWs or NPs, a large number of nanoparticles can also assemble near a NW forming multiple junctions. The possibility of forming a single junction can be increased by optimizing the concentration and volume of AuNPs and AgNWs. Figure (2b) shows transmission electron microscopy (TEM) and high-resolution TEM (HRTEM) images of a ~90 nm AuNP coupled to an AgNW of thickness ~ 100 nm. The HRTEM image shows a very small gap of 1.5-2 nm between the NW-NP junction. This single junction is achieved by initially dropcasting 1μL of $1.9 \times 10^5$ AgNWs on silicon wafer followed by dropcasting $1.5 \times 10^6$ AuNPs.

*(ii) Optical trapping*

The SPPs propagating along AgNW possess the gradient and scattering forces and thus can be used to trap the nanoparticles [30-33]. Figure (2c) shows a dynamical junction prepared by trapping the nanoparticle, using gradient force of SPPs and moved along NW by the scattering forces. Experiments reveal that upon excitation of plasmons along a ~15 μm long and ~ 300 nm thick AgNW with one end excitation using 1064 nm laser, a ~ 42 nm diameter $TiO_2$ particle moves towards the NW because of gradient force. Numerical simulations show that the scattering forces of SPPs push the particle away from the distal end, but experimentally, once the particle is trapped, it moves towards the excitation end and upon reaching the excitation end, gets pushed towards the surrounding medium. Particles with different sizes move with different velocities. Small particles have greater velocity than larger particles. Furthermore, the velocity of the particles with fixed size increases linearly with laser power.

The contradiction with theory suggests that there must be other forces acting on the $TiO_2$ nanoparticle. The laser excitation at the NW end and SPPs propagating along NW leads to the heating of fluid and thus a heat gradient exists in the system [34-36]. This creates a local thermal convection and leads to movement of particle from cold to hot region which is towards the excitation point [37,38]. The trapping and movement of nanoparticles were observed with a chamber of height 120 μm. Upon reducing the height of the chamber to 20 μm, the thermal convection is reduced [39] and no movement is observed. Thus, by changing the chamber height a dynamical or static junction can be prepared.

*(iii) Micromanipulation*

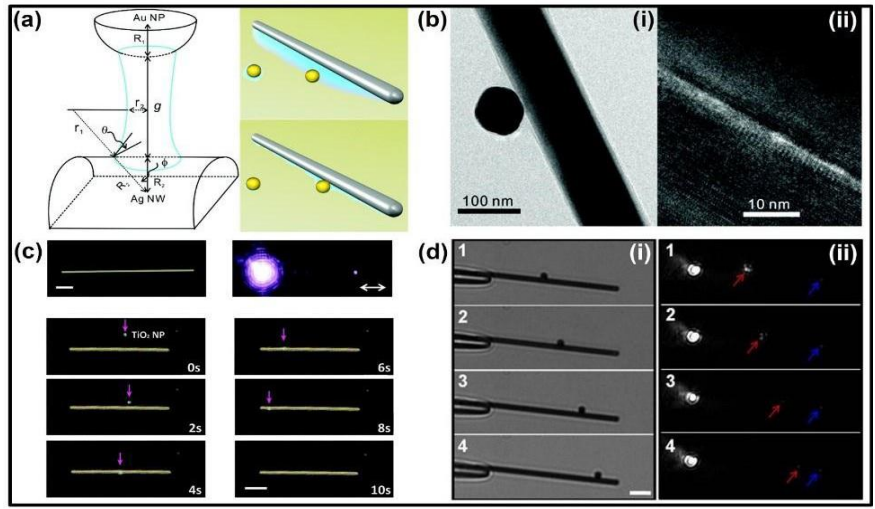

Fig 2. Production of nanowire-nanoparticle junctions by different methodologies. (a) Liquid-bridge between silver nanowire and gold nanoparticle dispersed in ethanol and representative images of a capillary force-induced self-assembled single silver nanowire-gold nanoparticle junction. (b) (i) and (ii) TEM and HRTEM images of a junction prepared via self-assembly process. Reproduced with permission from [27]. (c) Dynamic silver nanowire – $TiO_2$ junction. Surface plasmon polaritons (SPPs) were launched along silver nanowire with diameter ~300 nm by one end excitation with 1064 nm laser. The time series image of the movement of $TiO_2$ particle along silver nanowire supporting SPPs is shown. The scale bar is 5 μm. Reproduced with permission from [40]. (d) Manipulation of silver nanowire – silver nanoparticle junction using micro-manipulation of nanoparticle on the surface of nanowire using a tapered optical fiber. Left panel (i) shows the junction with different positions of the nanoparticle on nanowire and right panel (ii) shows the corresponding dark field images. Reproduced with permission from [40].

Fan *et al.* showed the formation of a dynamical junction where nanoparticle coupled with a silver nanowire can be moved along the length of silver nanowire by micromanipulation using optical fiber [40]. The position of NP along the NW has been shown to tune using a tapered optical fiber. A 785 nm laser was used to excite the SPPs along the NW. The position of a particle which is assembled near an AgNW can be tuned by continuously moving the particle or NW position using a 3D piezo stage. Figure (2d) shows manipulation of an AgNP along an AgNW imaged through a 50x 0.80 NA objective lens.

Figure 3 shows nanoparticle-nanowire junctions prepared using self-assembly technique. Figures (3a-b) show SEM and optical images of a junction formed with a silver nanowire of thickness ~100 nm and gold nanoparticle of diameter ~ 90 nm [27]. In this assembly, nanowire is of silver and nanoparticle is of gold, but both the structures can also be of same material. Figures (3c-d) show a SEM and optical image of a different junction made up of AgNW and silver nanoparticle [41]. The wire is 12 μm long and 250 nm thick and the diameter of the particle is 160 nm. Structures other than spherical particles can also be coupled to an AgNW for example, nanocubes and bipyramids [42-44]. For example in ref [45], silver nanocubes and bipyramid have been coupled to AgNWs using self-assembly.

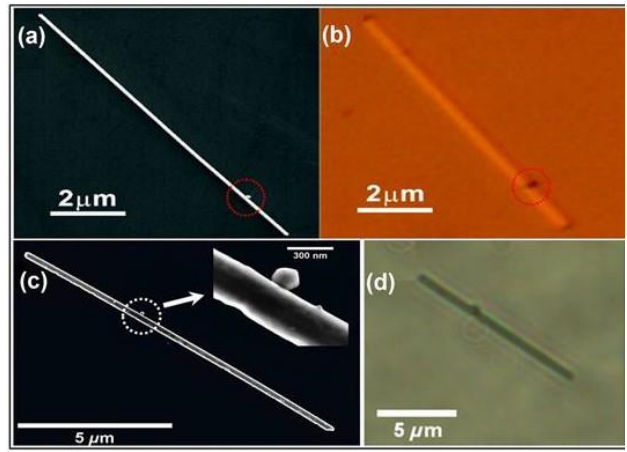

Fig 3. Various silver nanowire-nanoparticle junctions. (a) SEM image of silver nanowire- gold nanoparticle system prepared by capillary assisted self-assembly technique. (b) Corresponding optical image of the junction shown in (a). The red dotted circle indicates the junction. Reproduced with permission from [40]. (c) SEM image of silver nanowire- silver nanoparticle system also prepared by capillary assisted self-assembly technique. (d) Corresponding optical image of the junction shown in (c). The white dotted circle indicates the junction. Reproduced with permission from [41].

**4 Nanoparticle as an optial antenna: Coupling photons to plasmons**

A nanoparticle placed in the vicinity of a silver nanowire can act as an efficient antenna to couple the free photons into the propagating SPPs of AgNW [46]. The junction acts as a defect and the scattering of the light gives sufficient momentum to photons to couple to plasmons along the NW. Figure (4a) shows an oxidized junction created by coupling of silver nanoparticle to an AgNW. The NW is 5.96 μm in length and 259 nm thick and the attached NP is 660 nm in diameter. Upon excitation of NW the nanoparticle scatters the SPPs and converts them to free photons as shown in Fig (4b). Two bright spots are shown in the elastic scattering image other than the excitation spot. First one is the scattering from the nanoparticle and the other is from the distal end of NW. Upon excitation, the nanoparticle (Fig 4c), acts as an antenna and couples the photons to propagating plasmons. The SPPs propagating along NW, get scattered at the ends.

To study the polarization dependence of coupling of the photons to NW plasmons, an AgNP is attached to an AgNW with a bend or kink (Fig 4d). The point of NW-NP junction is marked as 'b' and the two ends of NW are marked as 'a' and 'd'. Point 'c' shows a kink in the NW which acts as a defect to scatter the SPPs. The NW is 16.65 μm long and 222 nm thick and the diameter of NP is 216 nm. Upon excitation of the junction with a 633 nm laser, the particle acts as an antenna and couples the free photons with propagating plasmons along the NW. The intensity of emission from the NW ends and also at the kink, can be tuned by changing the polarization at the junction. By changing the polarization of incoming laser at the junction, the intensity at the NW ends, 'a' and 'd', and also at the kink is shown to vary in Fig (4e). The coupling at the junction is maximum when the input polarization is along the junction, thus exciting the gap plasmons and minimum when exciting with polarization perpendicular to the junction [47,48].

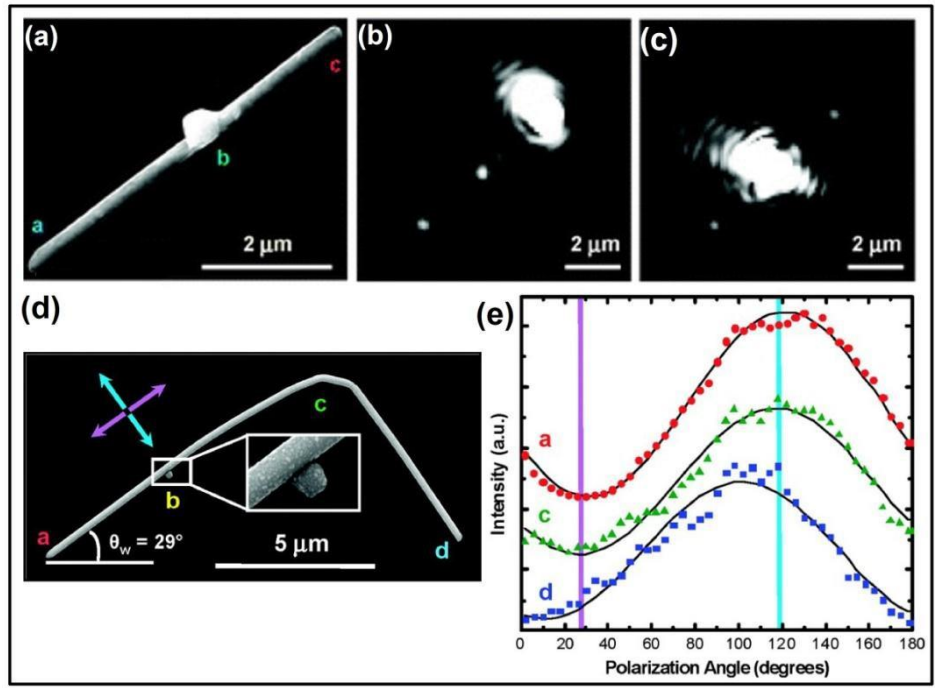

Fig 4. Nanoparticle as an optical antenna.**(a)** SEM image of a silver nanowire-nanoparticle junction after oxidation. **(b)** Coupling of light into the nanowire plasmons by excitation of the junction. **(c)** Coupling of light along silver nanowire through the excitation of nanowire end. **(d)** SEM image of a different silver nanowire-nanoparticle junction. The zoomed-in image of junction, marked as 'b' is shown in white rectangular box. **(e)** Output intensities from nanowire kink 'c' and both ends 'a' and 'd' as a function of polarization angle of input laser. Purple and light-blue lines show the polarization angle of the incident light along and perpendicular to the silver nanowire, respectively. Emission intensities are fitted using sine curves (black lines). Reproduced with permission from [46].

**5 Effect of nanoparticle on nanowire SPPs**

As explained in section 4, NP scatters the NW SPPs and the symmetry of SPPs propagating gets deformed [49,50]. The symmetric plasmonic field traveling along NW [51] gets distorted because of the non-uniform scattering by NP. Figure (5a) shows a 300 nm thick NW coated with a 50 nm layer of $Al_2O_3$, over which quantum dots were deposited. SPPs along the NW were launched by exciting one end of AgNW with a 633 nm laser source, with polarization along the NW length. These SPPs suffer scattering from NP placed in the vicinity of NW and get scattered. This scattering leads to the outcoupling of NW SPPs to free photons (Fig 5b). As shown in the fluorescent image (Fig 5c), the symmetric SPPs traveling along NW, after interaction with nanoparticle, get deformed and become zigzag in nature. This is because the symmetry of the structures is now broken and thus the field distribution now no longer remains symmetric.

SPPs power switching to different parts of branched NW by tuning the position of particle has also been shown using numerical calculations. The switch structure is shown in Fig (5d), with a NP of diameter 120 nm placed at a distance of 5 nm from a 120 nm thick AgNW with two branches at an angle of 60°. By changing the distance 'd' of NP from the kink, the input intensity $I_0$ can be switched in different branches of NW. Figure (5e) shows the tuning of output intensity $I_{out}$ in different branches of NW. At two different locations of NP on the NW, the majority of the SPPs field can be switched in either of the branch of NW as shown by numerical simulations Fig 5 (f,g).

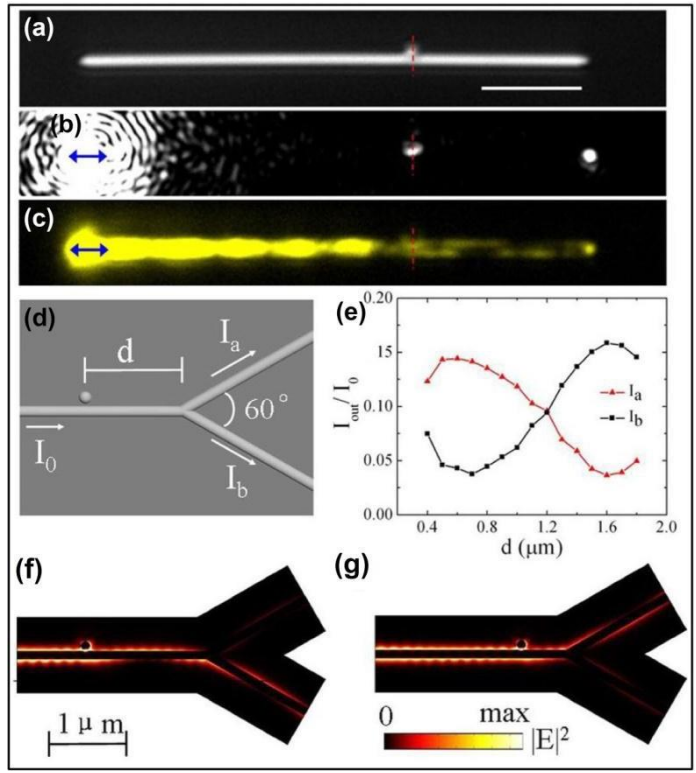

Fig 5. Scattering of nanowire SPPs by a nanoparticle. (a) Optical image of a silver nanowire–silver nanoparticle junction. (b) Elastic scattering image when SPPs are launched along the nanowire through the end excitation using 633 nm laser wavelength. The polarization (shown by the blue arrow) was kept along the nanowire length. (c) Quantum dots fluorescence image. The scale bar is 5 μm. (d) Schematic of a proposed switch structure. The nanowire and nanoparticle both are 120 nm in diameter and separated by 5 nm. (e) Intensity at the two branches normalized by input intensity as a function of nanoparticle position from the branches. (f,g) Electric field distribution along the proposed system for two different nanoparticle positions showing the power switch between two branches. Reproduced with permission from [49].

## 6 Harnessing large local electric field

### 6.1 Local excitation of surface enhanced Raman scattering

The junction created by coupling a NP to a NW creates a local large electric field and thus can be used for enhanced scattering processes. Molecules placed at the junction experience large electric field and thus undergo large spontaneous emission [52]. Figure (6b) shows a junction made up of gold NW and gold nanoparticle placed on a silicon substrate [53]. A thin layer of Malachite green isothiocynate (MGITC) molecules was dropcasted over the nanowires and nanoparticles. Although the molecules are strongly resonant at 633 nm wavelength, upon excitation of the molecules on the substrate no Raman signal was observed (black curve in Fig 6a).

Exciting the NW gives a weak SERS signal (red curve in Fig 6a). Upon excitation of the junction, a very intense SERS is observed because of the gap plasmons created at the junction (blue curve in Fig 6a). To study the input polarization dependence on the SERS intensity from the junction, a different junction was studied for two different input polarizations along and perpendicular to the junctions as shown in Fig (6c). SERS intensity from the junction is more intense for input polarization along the junction as compared to the polarization perpendicular to the junction. Numerical simulations performed at the junction for two different input polarizations show a large local electric field at the junction for polarization along the junction (Fig 6d) [54,55]. The calculated SERS enhancement at the junction for different input polarizations is also plotted which reveals that the maximum SERS enhancement is found when the polarization of incoming laser is along the junction [56,57]. The large local electric field produced at the junctions has been used in various detection and sensing devices [58-60]. A junction placed on a metal film supports more than one hotspot, because of the gaps created between NP (or NW) and metal film and have been used for enhancing emission from molecules and two dimensional materials placed near the junction.[61-63].

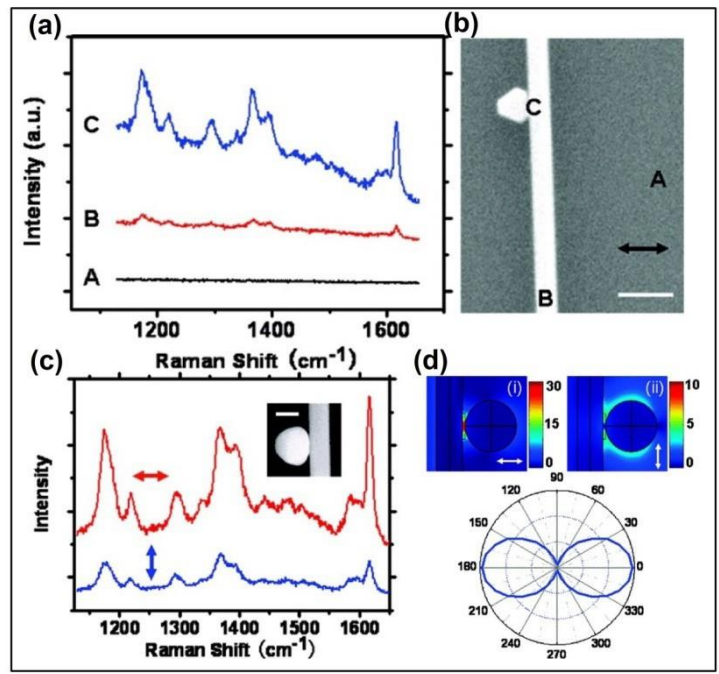

Fig 6. Nanoparticle-nanowire junction for polarization dependent surface enhanced Raman scattering (SERS). (a) SERS spectra of Malachite green isothiocyanate (MGITC) molecules at the position A, B and C described in SEM image. (b) of gold nanowire-nanoparticle junction. The scale bar is 400 nm and the white arrow shows the polarization of 633 nm input laser. (c) SERS spectra of MGITC molecules for two different input polarizations from gold nanowire-nanoparticle junctions shown in the inset with a scale bar of 200 nm. (d) Numerically calculated near-field electric field at the junction with nanowire and nanoparticle of diameter 100 nm having a gap of 5 nm between the two structures. Electric near field for input polarization transverse to the nanowire (i) and longitudinal to the axis of the nanowire (ii). The polar graph shows SERS enhancement at the junction as a function of input polarization angle. Reproduced with permission from [53].

6 (*ii*) *Polarization resolved surface enhanced Raman scattering*

A large local field is created at the NW-NP junction when it is excited by light with polarization along the junction. This is because of the generation of gap plasmon modes.[64]. Thus, the molecules confined in the gap formed between nanowire and nanoparticle will experience large electric field when the input polarization is along the junction [65]. But because of the plasmon hybridization, the output polarization of the emission will also be affected which should be probed to understand the coupling at the junction [16]. Figure (7a) shows the optical image of one such junction and the inset shows the SEM image [66]. The junction is created by coupling a Biphenyl-thiol (BPT) molecules coated gold NP of size ~ 200 nm to a NW of thickness ~ 300 nm using self-assembly methodology.

One such junction is excited by a 633 nm laser using a high numerical aperture objective lens. Figure (7b) shows a larger SERS intensity for input polarization transverse to the nanowire, as compared to the longitudinal input polarization because of the formation of gap plasmons. For a constant transverse input polarization, the output analyzed spectra are shown in Fig (7c). Majority of the SERS emission from the junction is polarized transverse to the nanowire. Thus, the gap plasmon modes formed at the junction also lead to an emission of highly polarized SERS intensity from the molecules confined between the nanowire and nanoparticle.

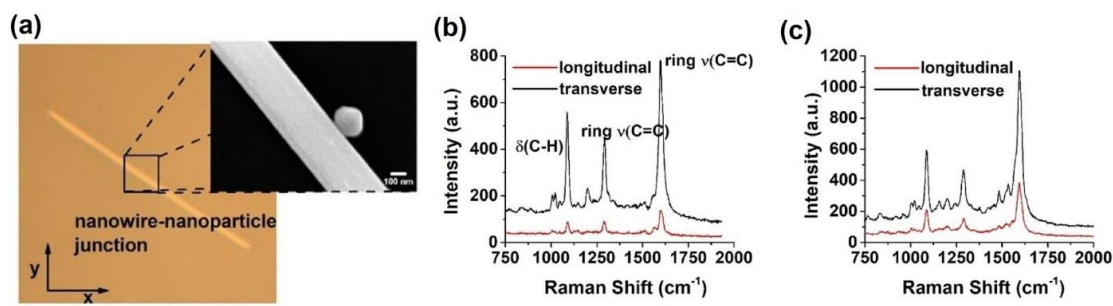

Fig 7. Input and output polarizations resolve surface enhanced Raman scattering (SERS) intensity from nanowire-nanoparticle junction. (a) Optical image of silver nanowire-gold nanoparticle junction. Inset shows the SEM image of the junction. (b) Input polarization dependence of SERS intensity from Biphenyl-thiol molecules confined in the gap between nanowire and nanoparticle. (c) Output polarization resolved SERS spectra for a fixed transverse input polarization. Reproduced with permission from [66].

6 (*iii*) *Remote excitation of surface enhanced Raman scattering*

*(a) Single path remote excitation*

The hotspot formed between an AgNW and AgNP creates a large local electric field because of gap plasmons formed between two metal surfaces. This junction facilitates a input polarization dependent large local electric field which has been shown to be used for enhanced spectroscopy studies. This junction has also been utilized for remote excitation of SERS from molecules confined in or near the junction [67- 69]. The propagating SPPs can be scattered by a NP placed near the vicinity of NW and this leads to a large local electric field because of enhanced scattering. Thus, the junction utilizes the field created by both the propagating SPPs and localized plasmons.

Figure 8 shows one such experimental configuration to utilize both the propagating SPPs of NW and localized plasmons of NP to achieve remote excitation of SERS from *para*-Aminothiophenol (*p*ATP) molecules confined in the nanogap [67]. Figures (8a) shows schematic representation for remotely exciting SERS. AgNPs were attached to a *p*ATP molecules coated AgNW. Focused excitation at one end of NW launches SPPs along the NW. The SPPs when interact with a nanoparticle create a hotspot and thus the molecules present at the gap experience a large electric field and undergo enhanced spontaneous emission.

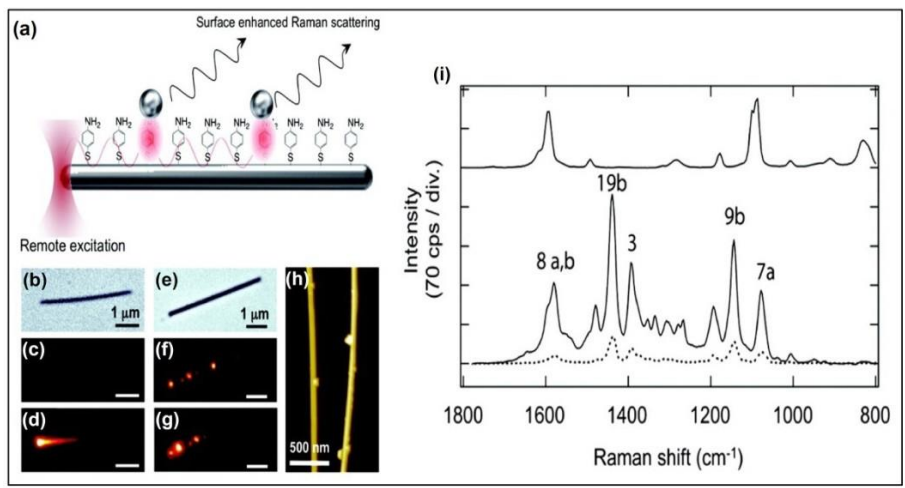

Fig 8. Remote excitation of surface enhanced Raman scattering (SERS). (a) Schematic representation of remote SERS at silver nanowire-nanoparticle junction. (b) Transmission image of silver nanowire on glass substrate. (c, d) Corresponding scattering images of nanowire under wide field illumination and focused laser illumination, respectively. (e, g) Transmission image, wide field illumination scattering image and focused laser beam scattering image from *para*-Aminothiophenol (*p*-ATP) molecules coated nanowire with several nanoparticles coupled to nanowire. (h) Atomic force microscopy (AFM) images of two *p*-ATP molecules coated silver nanowires with absorbed silver nanoparticles. (i) Raman spectra of *p*-ATP molecules (Top) and SERS from *p*-ATP molecules at the junction (Bottom) with two input polarizations. The black solid and dotted lines correspond to polarization along and transverse to the long axis of silver nanowire, respectively. Reproduced with permission from [67].

The optical images of *p*ATP coated NW show interesting differences before and after coupling NPs on NW. AgNWs of thickness 50-150 nm and length 2-30 μm were coated with *p*ATP molecules and were dropcasted on glass substrate. Because of the strong thiol bonding with silver the *p*-ATP molecules get attached to the AgNW [70]. Over this sample AgNP of average diameter 40 nm were spincoated. Figures (8b-d) show a transmission image of a *p*ATP molecules coated silver nanowire placed on a glass substrate, corresponding scattering image under wide field illumination and focused 633 nm laser image, respectively. In contrast to this, a *p*ATP coated AgNW with attached AgNP looks very different. Figures (8e-g) show a transmission image, wide field illumination charge and an image with focused 633 nm laser of NW, respectively. Both bright field image and focused laser at one end of AgNW show multiple spots on the NW, because of plasmon to photon conversion at the junction. Atomic force microscopy (AFM) image of a *p*ATP coated AgNW with two NPs is also shown in Fig (8h) which shows adsorption of two individual NPs on AgNW. Aggregates of NPs on AgNW were also reported. One such junction with *p*ATP coated AgNW of 10 μm with attached NPs, was used for remote SERS studies. The bright spot at the junction is around 7 μm away from the excitation spot. Figure (8i) shows a remotely collected SERS spectrum (bottom) with two input polarizations, along the wire (black solid) and transverse to NW (black dotted). For comparison, Raman spectra of *p*ATP molecules is given in Fig (8 i) (top spectrum)..

*(b) Dual path remote excitation*

Since SPPs along the nanowire can be launched by exciting either of the ends of NW, both the ends can be excited simultaneously which can lead to plasmon hybridization [41,71]. This hybridization of plasmons creates maxima and minima of electric field throughout the NW. A 'hotspot' of electric field can be created by placing a particle at the maxima of electric field along a silver nanowire having counter-propagating plasmons. These hotspots have been utilised for enhanced spectroscopy.

Figure 9 shows one such study of surface enhanced Raman scattering (SERS) from Nile blue molecules placed near the junction [41]. AgNW and AgNP suspension was mixed with 5 μM Nile blue molecules and were dropcasted on glass substrate. The inset in Fig 9 shows the SEM image of the junction. The NW is 12 μm long and 250 nm thick and the diameter of NP is 160 nm. For the dual excitation, a 633 nm laser of power 2 mW was split into two beams with power 1 mW in each path. In single path configuration, one end of AgNW was excited by a 633 nm laser with polarization along the length of NW. Figure (9 a, b) shows single path remote excitation of junction by exciting End 1 and End 2, respectively. The black and the red curve show the SERS intensities from the junction upon the excitation of End 1 and End 2 of the silver nanowire, respectively. When both the ends are excited simultaneously a much stronger SERS signal is obtained which is more than the sum of two individual SERS signals obtained in single path excitation. In another study with this structure, the polarization of the two incoming beams has also been changed one by one, keeping constant the other end polarization and the strong modulation of SERS intensity signal is reported.

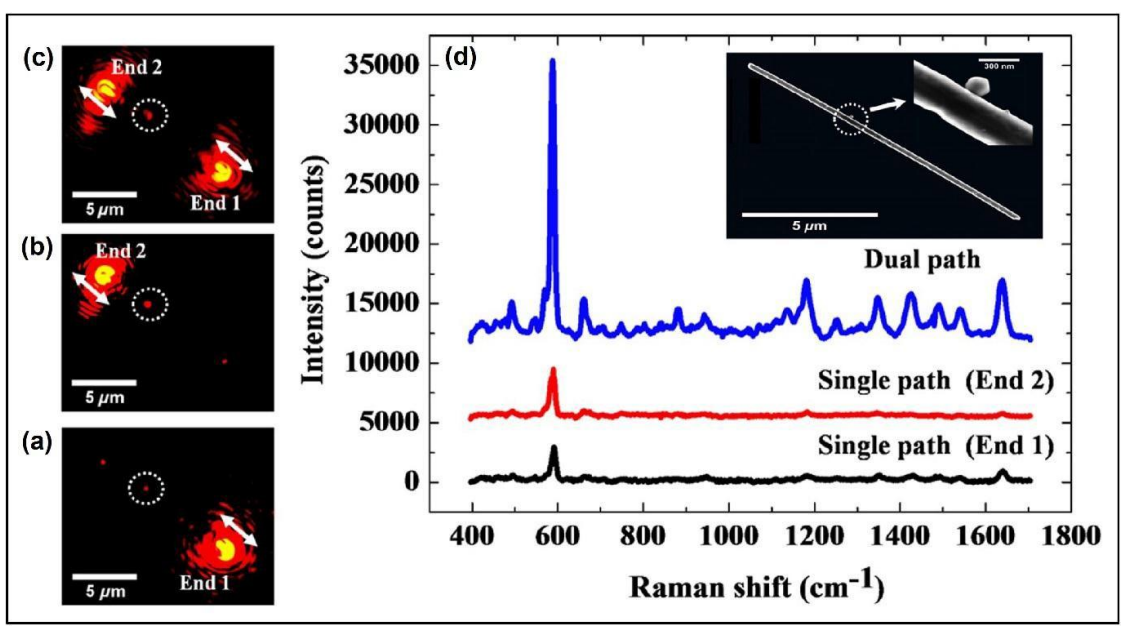

Fig 9. Remote excitation surface enhanced Raman scattering (SERS) in dual path excitation. (a) Elastic scattering image of silver nanowire-nanoparticle junction upon illuminating End 1. Arrow shows the polarization of the input laser beam. White dotted circle shows the emission from the junction. (b, c) Elastic scattering image of the junction upon excitation of End 2 and both the ends, respectively. (d) Nile blue SERS spectra collected from the junction in different excitation schemes. Inset shows SEM image of junction. SERS spectrum when End 1 is excited (black curve). SERS spectrum when End 2 is excited (red curve). SERS spectra when both the ends are excited simultaneously (blue curve). Reproduced with permission from [41].

**7 Nanowire-nanoparticle junction as a directional antenna**

Optical antennas can be intermediates between far-field radiation and the near-fields around quantum emitters such as atoms, molecules and quantum dots [72]. Understanding the field enhancement capabilities and directivities of optical antennas in the form of plasmonic nanostructure has turned out to be an important task. To this end, various optical antenna effects for elastic and inelastic light scattering have been explored [73-76], and below we give a brief overview of some relevant topics.

*7. (i) Directly excited antenna*

The wavevectors of emission from the molecules confined in the junction can give information about the orientation of the molecules inside the junction and also about the interaction of molecules [77-79]. The wavevectors can also provide the information about the plasmon modes formed between the NW and the junction. The wavevectors of emission can be probed by Fourier plane imaging, which quantifies the angular spread of the emission. Fourier plane imaging images the back focal plane of the objective lens used for the collection of the emission. The emission is quantified into radial and azimuthal angles, where the radial angles are limited by the numerical aperture of the objective lens used [80-82].

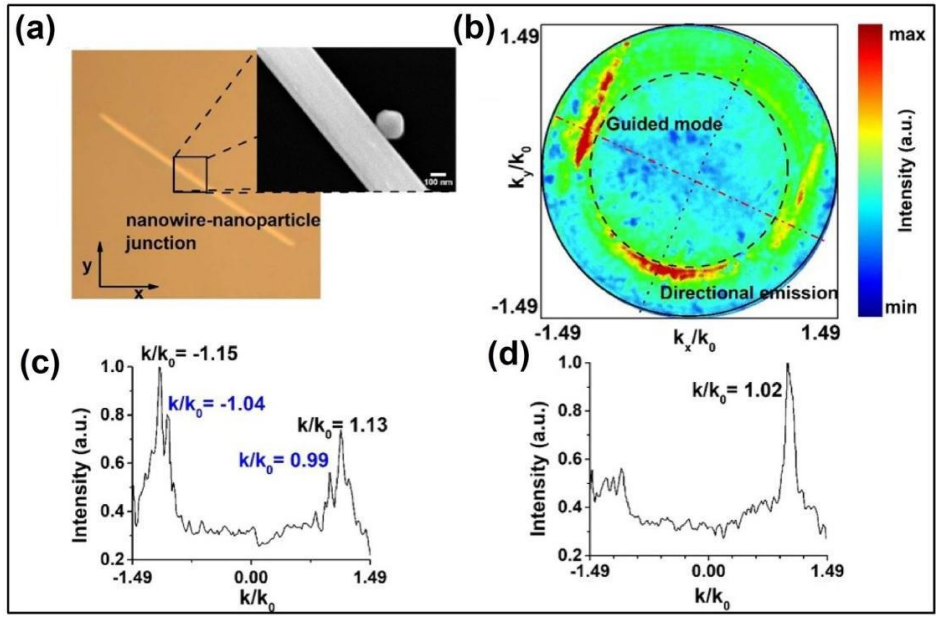

Fig 10. Fourier plane imaging of SERS emission from BPT molecules confined in the gap between silver nanowire-gold nanoparticle. (a) Optical image of silver nanowire-gold nanoparticle junction. Inset shows the SEM image of the junction. (b) Corresponding Fourier plane image of the SERS emission after rejecting 633 nm laser light used for exciting the junction. (c) Intensity distribution along the red dotted line in (b), showing the wavevector emission for guided modes of nanowire.(d) Intensity cross-cut along the black dotted line in (10b), showing the showing the angular spreading of SERS emission wavevectors from the junction. Reproduced with permission from [66].

Figure 10 shows Fourier plane imaging of SERS of Biphenyl-thiol (BPT) molecules confined in the gap [66]. The optical image of the junction is shown in Fig (10a) and the inset shows the SEM image of the junction. The junction is created by coupling a BPT molecules coated gold NP of size ~200 nm to a NW of thickness ~ 300 nm using self-assembly methodology. A high numerical aperture lens (100×, 1.49 NA) was used to excite the junction with 633 nm laser. The SERS spectrum is shown in Fig (7b). For Fourier plane imaging, the polarization of the input light is along the junction. Figure (10 b) shows the Fourier plane captured by rejecting the excitation wavelength. The emission from the junction is directed to a narrow range of radial angles around $k_x/k_0 = 1.02$ which is near to the glass-air interface. The full width of half maxima of the emission is $\Delta (k/k_0) = 0.21$. The emission pattern also shows two leaky modes of SPPs traveling along the NW in two opposite directions perpendicular to NW axis. The SERS photons because of the near field coupling get coupled to plasmons in the wire in different modes. The wavevectors of the leaky modes which are observed in the Fourier plane images are $k/k_0 = -1.04; 0.99$ and $-1.15; 1.13$. This study also shows that as the thickness of the NW is reduced, the emission wavevectors cover a broader range of wavevectors with no signatures of leaky modes, as the number of leaky modes supported by NW reduces as with a reduction in the NW thickness [83,84].

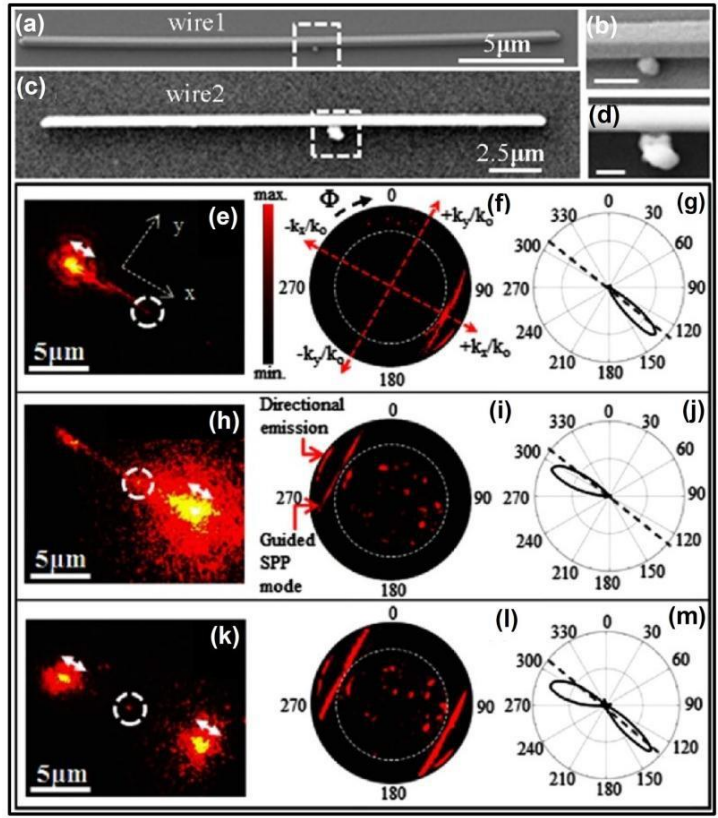

Fig 11. Fourier plane imaging of scattered surface plasmonpolaritons of silver nanowire from a silver nanoparticle. (a,c) SEM images of silver nanowire - silver nanoparticle junctions. (b,d) Magnified image of the junctions shown in (a,c), respectively. (e) Elastic scattering image of the junction shown in (a) when one end of the nanowire is illuminated with 633 nm wavelength through the glass side. Arrow indicates the input polarization. The dashed white circle shows the emission from the junction. (f) Fourier plane image of the emission from the junction. (g) Intensity distribution along the azimuthal angle (white dotted circle) at a constant radial angle in the (f). (h, j) Scattering image, Fourier plane image and intensity distribution of the emission from the junction when exciting the other end of the nanowire with an air immersion lens. (k, m) Scattering image, Fourier plane image and intensity distribution of the emission from the junction with both the excitations. Guided SPPs modes near glass air critical angle and directional emission at supercritical angles are shown in (i). The collection is always through the glass side. Reproduced with permission from Ref [85].

*7 (ii) Remotely excited antenna*

What are the wavevectors of scattered SPPs propagating along a NW by a NP placed in the vicinity of NW? To answer this question, Singh *et al.* have performed Fourier plane imaging on emission from the junction [85]. A nanowire-nanoparticle junction was prepared using self-assembly methodology by dropcasting AgNW and AgNP together and left to dry. Figure (11a) shows the SEM image of one such sample. The magnified image of the junction shows that a single NP is sitting in the vicinity of a NW placed on a glass substrate. The configuration used for launching the plasmons along the NW is dual excitation. The NW end can be excited either using an air immersion objective lens (100x, 0.90 NA) through the air or using an oil immersion objective lens (60×, 1.42 NA) through the glass. In this study three configurations have been used which are following:

(i) Single path excitation through air immersion objective lens and glass side collection with oil immersion objective lens.

(ii) Single path excitation and collection through oil immersion objective lens.

(iii) Dual path excitation through air immersion and oil immersion objective lens and glass side collection with oil immersion objective lens.

In the first configuration, one end of AgNW was excited through the glass side and the SPPs along the wire get scattered by the NP (Fig 11e). The scattered photons from the junction were collected using the oil immersion objective lens to perform Fourier plane imaging. The emission profile from the junction shows two interesting features (Fig 11 f). One, the guided modes of SPPs propagating along the NW with emission angles around the critical angle of glass-air interface. Another feature is the directional emission along the NW longitudinal axis at the supercritical angles. The emission angles are quantified by measuring the azimuthal spreading and is found to be 21° (Fig 11 g). With the excitation through the air side (Fig 11h), the junction was spatially filtered and the scattering from the junction is again directional in nature with an azimuthal spreading of 24° (Fig 11 i,j). With both the excitations, in counter propagating SPPs modes, the emission profile showed two guided modes and two arcs at supercritical angles showing directional emission because of plasmon scattering by nanoparticle.

*8* **Coupling quantum objects to nanowire**

Optical information processing using quantum mechanical effects open new avenues in quantum optics and quantum information processing [86-89]. One of the prototypical systems that enable the testing of quantum optical effect at subwavelength scale is a semiconducting artificial atom called the quantum dot. Given that these structures can be chemically produced with great control in terms of size, shape, stability and band-gap tunability, quantum dots have emerged as excellent tools for solid-state quantum optics [90, 91]. By interfacing such quantum systems to plasmonic nanowires, single plasmon generation in nanowires have been shown [92]. Also the coupling of quantum objects to nanowires can lead to hybrid couplings that facilitate optical resonances along with localized and delocalized electric fields [93-96]. The field created by the interference of propagating SPPs of AgNW and localized plasmons of AgNW can be used to modulate the emission intensity from a single quantum dot placed in the vicinity of NW. One end of the NW can be used to excite propagating SPPs along NW and by focusing one more laser at the center of NW, localized surface plasmons can be excited. A single quantum dot can be placed at the point of interference of two types of plasmonic fields and depending on the phase difference between the two plasmonic fields the intensity at the junction can be modulated [97].

Figure (12a) shows the schematic of experimental configuration for modulating the fluorescence emission of a single quantum dot placed near the vicinity of a single NW. The propagating SPPs along NW are excited by a focused laser beam, EXC 1. This field interacts with the localized plasmons generated by focusing another laser beam, EXC 2, at the place where a single quantum dot is placed near an AgNW of diameter 84 nm and length 12.8 μm. Optical transmission and AFM images of AgNW are shown in Fig (12b). A 15 nm layer of $Al_2O_3$ is coated on the AgNW to prevent the quenching of quantum dot fluorescence.

EXC 1 at the end of AgNW launches propagating plasmons along the NW, whereas EXC 2 at quantum dot results in localized plasmons at the junction. The polarization of EXC 1 is kept along the length of NW and of EXC 2 is kept perpendicular to the NW to create a large local electric field at the junction where the two fields are interacting. The scattering image of the system is shown in Fig (12c) with dual excitation. One of the excitation is at the NW end (shown with rectangular box B) and the other excitation is at quantum dot (shown with rectangular box A). The other end of AgNW (C) shows outcoupling of plasmons to freely propagating photons. The fluorescent image in Figure (12d) shows three bright spots, where A is the emission at the location of quantum dot attached to NW. The emission from the quantum dot at the location A is getting coupled to the plasmons in the NW. The emission at the scattering points end 1 and end 2 of NW are shown as points B and C in Fig (12d). Figure (12e) shows the time trace of quantum dot fluorescence intensity. The intensity is recorded at 10 frames per second. The time trace shows the blinking of fluorescence intensity from the quantum dots and thus it confirms the quantum dot to be a single photon source [98]. The other two fluorescence emission from the NW ends also show the blinking behavior. By varying the phase between EXC1 and EXC2 using Soleil-Babinet compensator, quantum dot fluorescence intensity has been tuned from a minimum value of 0 to 12300 counts (Fig 12f). The blue curve in (f) shows the intensity from the quantum dot in the absence of EXC 1. The intensity counts are 3200 counts. To achieve the same intensity from the quantum dot in two excitations, EXC 1 and EXC 2, the input intensity of EXC 1 is taken to be 4.5 times more than EXC 2.

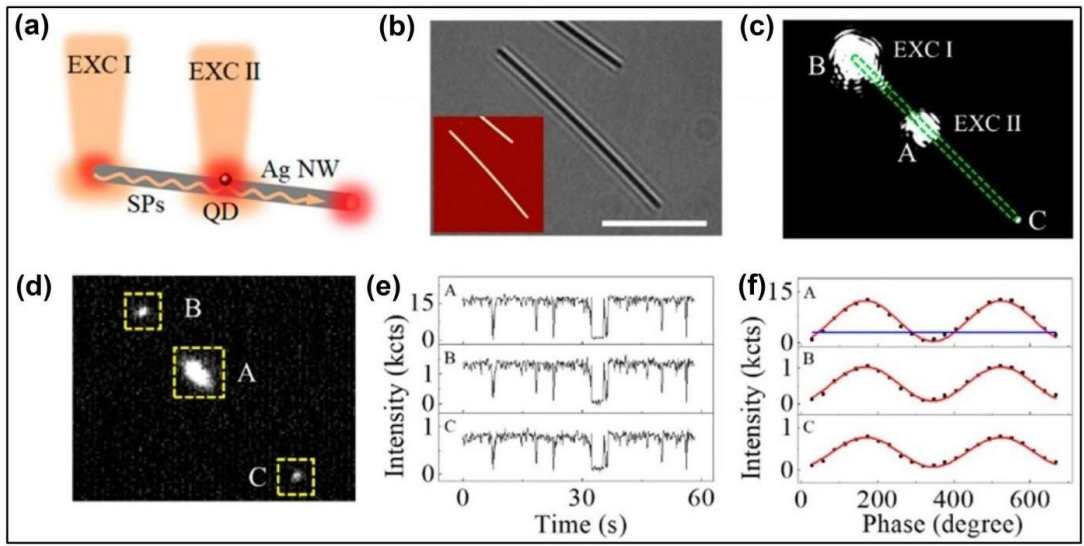

Fig 12. Modulated quantum dot emission using interference between propagating SPPs and localized plasmons. (a) Schematic of the hybrid single quantum dot - silver nanowire system. The system is excited by two coherent laser beams, excitation 1 (EXC I) at the end of the nanowire and excitation 2 (EXC II) at the quantum dot. The excited SPPs are represented by wavy orange line. (b) Optical image of the QD-NW system under study. Inset shows the AFM image of the system. The scale bar is 5 μm. (c) Scattering image of the QD-NW system under two excitations. Green dashed lines are the outline of the nanowire. (d) Fluorescence image of the same QD-NW system. The bright spots at both the ends of the nanowires (B and C) show the scattering of the fluorescence emission of the quantum dot (from point A) coupled into the nanowire. (e) Emission intensities at the points A, B and C as a function of time recorded at 10 frames per second. The intensity unit kcts stands for 1000 counts. (f) Emission intensities at the points A, B and C with the change in the phase of the laser beam at the EXC I. The blue line in the plot at point A shows the emission at A without the presence of EXC II. Points are fitted using sine curves. Reproduced with permission from [97].

This is because the propagating SPPs and localised SPPs produce different electric field intensity at the position of quantum dot. The output from both the ends of NW also show the same modulation of the intensity depending on the phase between the two excitations. The oscillation period of modulated emission from quantum dot and NW ends has a period of $2\pi$.

*9* **Emerging prospects**

Now that we have got an overview of how plasmonic nanowire-nanoparticle junction can be prepared and utilized for various light-matter interaction studies, we conclude by noting certain emerging research directions and questions that utilize the said geometry.

- *Strong coupling physics and hybridization of optical resonances*: Strong coupling physics and its applications have evolved into an important area of nanophotonics. At the heart of this study is the coupling of modes between two different resonators and how they strongly interact to create hybridized states. An interesting prospect to explore is how to strongly couple the delocalized plasmons to nanophotonic emitters including single molecules, quantum dots and emissive nanoparticles.
- *Explore the quantum regime*: Controlling the chirality, phase and intensity of quantum optical elements have emerged as an important task of quantum nanophotonics. To this end, plasmon-polaritonic junction like nanowire-nanoparticle junction can introduce interesting physics in the form of spin-momentum locking and Hall-effect of light. A lot of research is expected to come based on this concept, and the discussed geometry can play a vital role in this.
- *Nano-optoelectronic devices*: Wire is a fundamental element of any circuitry. By applying an electric bias to a nanophotonic wire, one can add a new parameter to nanophotonic excitation, where electric bias can be used to excite and control the plasmon propagation characteristics. Such attempts to control plasmons with electric excitation is already underway, and by employing such strategies to nanowire-nanoparticle junction, one can potentially introduce gating of optical processes, where the plasmonic hot spot can be tailored as a gate to control the intensity, phase, polarization and nonlinearity of optical interaction.
- *Opto-thermal devices*: Metals dissipate heat due to Ohmic losses. This is mainly governed by the absorption of light, and connected to the imaginary part of the dielectric constant. By designing and fabricating local perturbations and inclusions on the nanowires, heat propagation and localization can be controlled. Such structures can also have implications in nanophotonics.
- *Plasmon-complex fluid interaction*: Optical anisotropy is an important concept for display devices such as LCDs. Understanding how interfacing plasmons with soft-matter systems such as liquid crystals, colloids, polymers and vesicles can lead to some new insights on optical binding and optically induced forces at nanoscale. Moreover, such studies may also lead to some applications in biology where plasmons can be harnessed to provide confined light and heat sources.

Thus, we foresee interesting prospects of the nanowire-nanoparticle systems as versatile platform not only to study fundamental light-matter interaction studies at nanoscale, but also as a promising structure for various applications including on-chip optical processing (both at the classical and quantum levels), single molecule spectroscopy, nonlinear nanophotonics and nanoscale opto-thermal physics.

**Acknowledgements**

This work was partially funded by Air Force Research Laboratory and DST Energy Science (SR/NM/TP-13/2016) grant. ST and CT thank Utkarsh Khandelwal, Suryanarayan Banerjee, Vandana Sharma for fruitful discussions. S.T. thanks Infosys foundation and C.T. thanks Inspire fellowship for financial support. GVPK acknowledges DST for Swarnajayanti fellowship.